\documentclass[]{article}

\usepackage{amsfonts,amsmath,amssymb}
\usepackage{amsthm}
\usepackage[cp866]{inputenc}
\usepackage[english]{babel}

\allowdisplaybreaks \theoremstyle{definition}
\newtheorem{Definition}{Definition}

\theoremstyle{plain}
\newtheorem{Theorem}{Theorem}
\newtheorem{Corollary}{Corollary}
\newtheorem{Proposition}{Proposition}

\newtheorem{Lemma}{Lemma}

\newcommand{\bfSig}{\mathbf{\Sigma}}

\title{Effective Wadge Hierarchy\\ in Computable Quasi-Polish Spaces}
\author{Victor Selivanov\thanks{This research was partially supported by  RFBR-JSPS Grant 20-51-50001.}
\\A.P. Ershov
Institute of
Informatics Systems SB RAS\\ 
{\tt vseliv@iis.nsk.su}
}

\begin{document}
\large
\date{}
 \maketitle

\begin{abstract}
We define and study an effective version of the Wadge hierarchy in  computable quasi-Polish spaces which include most spaces of interest for computable analysis. Along with hierarchies of sets we  study  hierarchies of $k$-partitions which are interesting on their own. We show that levels of such hierarchies are preserved by the computable effectively open surjections, that if the effective Hausdorff-Kuratowski theorem holds in the Baire space then it holds in every computable quasi-Polish space, and we extend the effective Hausdorff theorem to $k$-partitions.


%{\bf MSC}: Primary   03D15, 03D78, 58J45; Secondary 65M06, 65M25.
\end{abstract}

\section{Introduction}\label{in}

Classical  descriptive set theory (DST) \cite{ke95} deals with
hierarchies of sets, functions and equivalence relations in Polish
spaces.  Recently, classical DST was extended to quasi-Polish spaces which contain many important non-Hausdorff spaces \cite{br}.
Theoretical  Computer Science and Computable Analysis   especially need an effective  DST for some effective versions of the mentioned
spaces. A lot of useful work in this direction was done in Computability Theory but mostly for very special spaces like the discrete space $\mathbb{N}$, the Baire space $\mathcal{N}$, and some of their  relatives \cite{ro67,mo09}. 
Effective versions of classical Borel, Hausdorff and Luzin hierarchies are naturally defined for every effective space (see e.g.\cite{s06}) but, as also in the classical case, they behave well only for spaces of special kinds. Recently, a convincing version of a computable quasi-Polish space (CQP-space for short) was suggested in \cite{br1,hs}.

In this note we continue to develop effective DST in CQP-spaces  where effective analogues of some important properties of the classical hierarchies hold. Namely, we develop an effective Wadge hierarchy (including the hierarchy of $k$-partitions) in such spaces which subsumes the effective Borel and  Hausdorff hierarchies (as well as many others) and is in a sense the finest possible hierarchy of effective Borel sets. In particular, we show that levels of such hierarchies are preserved by the computable effectively open surjections, that if the effective Hausdorff-Kuratowski theorem holds in the Baire space then it also holds in every CQP-space,  and we extend the effective Hausdorff theorem for CQP-spaces from \cite{s15} to $k$-partitions. 

Technically, the effective Wadge hierarchy suggested here is an instantiation  of the so called fine hierarchy (FH) introduced and studied in a series of my publications (see e.g. \cite{s08} for a survey). Modulo these publication, proofs here are short and technically easy, so the definition and nice properties of the effective Wadge hierarchy are perhaps a major contribution of this paper because it brings the investigation of effective hierarchies in CQP-spaces to the stage of maturity. 

The preservation property of the effective Wadge hierarchy established in this paper seems to be a principal property of the effective hierarchies which has several applications, for instance to effective Hausdorff-Kuratowski-type theorems and to the non-collapse properties of the effective hierarchies; we demonstrate this by corresponding examples. The proof of the  preservation property  is based on the observation in \cite{ch} which effectivises the corresponding classical fact \cite{sr07,br} and shows that Baire category technique is consistent with effectivity. 

The computable effective open surjections which are principal in our paper are also important in Computable Analysis because they satisfy effective version of some classical theorems on continuous open surjections (see e.g. \cite{he99,zi06} and references therein).   The preservation property  means that the effective Wadge hierarchy (hence also the effective versions of classical hierarchies) behaves well in the computable effective open images of many spaces of interest to core mathematics.

Already the ``correct'' extension of the Wadge hierarchy to arbitrary quasi-Polish (and even Polish) spaces is far from obvious because the structure of Wadge degrees in non-zero-dimensional such spaces is usually very complicated, hence can not be used to identify the Wadge hierarchy in this structure as it is done for the Wadge hierarchy in zero-dimensional spaces (see the recent preprint \cite{s19} for a detailed discussion). The ``correct'' definition of the effective Wadge hierarchy is still more problematic because the structure of effective Wadge degrees is complicated already in zero-dimensional spaces (and even in the discrete space of natural numbers were it coincides with the structure of many-one degrees). To my knowledge, our approach to define the  FHs via set operations used in this paper is currently the only available approach to identifying the effective Wadge hierarchy.

We start  in the next section with recalling definitions of relevant notions  of the effective DST. In Section \ref{fhkpart} we discuss relevant technical notions and facts about the FH. In Section \ref{wkpart} we introduce the effective Wadge hierarchy  and establish some of its basic properties. In Section \ref{preserve} we establish the preservation property. In Section \ref{trans}  we discuss transfinite extensions of the effective hierarchies which are  used in Section \ref{efHau} to investigate the effective Hausdorff-Kuratowski theorem for $k$-partitions. We conclude in Section \ref{con} with mentioning some open questions.

\section{Preliminaries}\label{prel}

In this section we briefly recall some notation, notions and facts used throughout the paper. Some more special information is recalled in the corresponding sections below. 

\subsection{Spaces and trees}\label{efclass}

We use standard set-theoretical notation. In particular, $Y^X$ is the set of functions from $X$ to $Y$,
$P(X)$ is the class of subsets of a set $X$, $\check{\mathcal{C}}$ is the class of complements $X\setminus C$ of sets $C$ in $\mathcal{C}\subseteq P(X)$, and $BC(\mathcal{C})$ is the Boolean closure of $\mathcal{C}$.

All considered spaces  are assumed to be countably based $T_0$ (sometimes we call such spaces cb$_0$-spaces). 
By {\em  effectivization of a cb$_0$-space $X$} we mean a numbering $\beta:\omega\to P(X)$ of a base in $X$ such that there is a uniform sequence $\{A_{ij}\}$ of c.e. sets with
$\beta_i\cap\beta_ j=\bigcup\beta(A_{ij})$, where $\beta(A)$ is the image of $A$ under $\beta$. The numbering $\beta$ is called an {\em effective base of $X$} while the pair $(X,\beta)$ is called an  effective space. We simplify $(X,\beta)$ to $X$ if $\beta$ is clear from the context.  \emph{Effectively open sets} in $X$ are the sets of the form $\bigcup_{i\in W}\beta(i)=\bigcup\beta(W)$, for some c.e.~set~$W\subseteq\mathbb{N}$. The standard numbering $\{W_n\}$ of c.e. sets \cite{ro67} induces a numbering of the effectively open sets. The notion of effective space allows to define e.g. computable and effectively open functions between such spaces \cite{wei00}. A function $f:(X,\beta)\to(Y,\gamma)$ is {\em computable} if  $f^{-1}(\gamma_n)=\bigcup\beta(W_{g(n)})$ for some computable function $g:\mathbb{N}\to\mathbb{N}$. A function $f:(X,\beta)\to(Y,\gamma)$ is {\em effectively open} if  $f(\beta_n)=\bigcup\gamma(W_{h(n)})$ for some computable function $h:\mathbb{N}\to\mathbb{N}$.

Many popular spaces (e.g., the discrete space $\mathbb{N}$ of natural numbers, the space ${\mathbb R}$ of reals, the Scott domain $P\omega$ of all subsets of $\mathbb{N}$, the Baire space $\mathcal{N}=\mathbb{N}^\mathbb{N}$, the Cantor space  $A^\omega$ of infinite words in a finite alphabet $A$) are effective spaces (with natural numberings of bases). The effective  space $\mathbb{N}$ is trivial  topologically but very interesting for Computability Theory. 

We use standard notation related to the Baire space. In particular, $\omega^*$ is the set of finite strings of natural numbers including the empty string $\varepsilon$, $|\sigma|$ is the length of a string $\sigma$, $[\sigma]$ is the basic open set induced by $\sigma\in\omega^*$ consisting of all $p\in\mathcal{N}$ having $\sigma$ as a prefix.  By a {\em tree} we mean a nonempty initial segment of $(\omega^*;\sqsubseteq)$ where $\sqsubseteq$ is the prefix relation. An  {\em infinite path} through a tree $T$ is an element $p\in\mathcal{N}$ such that $p[n]\in T$ for each $n$ where $p[n]$ is the prefix of $p$ of length $n$. A tree $T$ is {\em well founded} if there is no infinite path through $T$.

\subsection{Effective versions of classical hierarchies}\label{efhier}

Let  $\{\Sigma^0_{1+n}(X)\}_{n<\omega}$ be the effective Borel hierarchy and  $\{D_n(\Sigma^0_m(X))\}_n$ be the effective Hausdorff difference hierarchy over $\Sigma^0_m(X)$  in arbitrary effective space $X$. Another popular notation for levels of the difference hierarchy is $\Sigma^{-1,m}_n=D_n(\Sigma^0_m(X))$, with $\Sigma^{-1,1}$ usually simplified to $\Sigma^{-1}$.  Let also $\{\Sigma^1_{1+n}(X)\}$ be the effective Luzin hierarchy. We do not repeat the standard definitions but mention that they come together with standard numberings of all levels of the hierarchies, so we can speak e.g. about uniform sequences of sets in a given level. We use definitions based on set operations, they may be found e.g. in \cite{s06,s15}; there is also an equivalent approach based on the Borel codes \cite{lo78,ke95}.  E.g., $\Sigma^0_1(X)$ is the class of effectively open sets in $X$,   $\Sigma^{-1}_2(X)$ is the class of differences of $\Sigma^0_1(X)$-sets, and $\Sigma^0_2(X)$ is the class of effective countable unions of $\Sigma^{-1}_2(X)$-sets.  

Levels
of the effective hierarchies are denoted in the same manner as levels of the
corresponding classical hierarchies, using the lightface letters
$\Sigma,\Pi,\Delta$ instead of the boldface
$\bf{\Sigma},\bf{\Pi},\bf{\Delta}$ used for the classical
hierarchies \cite{ke95}. The boldface classes may be considered as ``limits''  of the corresponding lightface levels (where the limit is obtained by taking the union of the corresponding relativised lightface levels, for all oracles). Thus the effective hierarchies not only refine but also generalise the classical ones.

\subsection{Effective versions of the Wadge hierarchy}\label{efwadge}

Here we discuss the effective Wadge hierarchy (EWH) informally, in order to give the reader a first impression on how it looks like. Precise definitions are rather technical, so we postpone presenting them to Section \ref{fhkpart}.

By the EWH in a given effective space we mean a special case of the so called fine hierarchy (FH) introduced and studied in a series of my publications (see e.g. \cite{s08} for a survey). We briefly recall some relevant notions and properties of the FH.
By a {\em base in a set $X$} we mean a sequence $\mathcal{L}=\{\mathcal{L}_n\}_{n<\omega}$ of subsets of $P(X)$ such that any $\mathcal{L}_n$ is closed under union and intersection, contains $\emptyset,X$ and satisfies $\mathcal{L}_n\cup\check{\mathcal{L}}_n\subseteq\mathcal{L}_{n+1}$. For this paper, the {\em effective Borel bases} $\mathcal{L}(X)=\{\Sigma^0_{1+n}(X)\}$ in  effective spaces $X$ are especially relevant. 

The FH over the base $\mathcal{L}$ is a sequence $\{\mathcal{S}_\alpha\}_{\alpha<\varepsilon_0}$, $\varepsilon_0=sup\{\omega,\omega^\omega,\omega^{\omega^\omega},\dots\}$, of subsets of $P(X)$ constructed from the sets in levels of the base by induction on $\alpha$ using suitable set operations. 
The construction is designed in such a way that $\mathcal{S}_\alpha\cup\check{\mathcal{S}}_\alpha\subseteq\mathcal{S}_\beta$ for all $\alpha<\beta<\varepsilon_0$, and the FH subsumes many refinements of the base $\mathcal{L}$. In particular $\mathcal{L}_0=\mathcal{S}_1$, $\mathcal{L}_1=\mathcal{S}_\omega$, $\mathcal{L}_2=\mathcal{S}_{\omega^\omega}$,  $\mathcal{L}_3=\mathcal{S}_{\omega^{\omega^\omega}},\dots$, $\{\mathcal{S}_n\}_{n<\omega}$ is the difference hierarchy over $\mathcal{L}_0$, $\{\mathcal{S}_{\omega^{n+1}}\}_{n<\omega}$ is the difference hierarchy over $\mathcal{L}_1$, $\{\mathcal{S}_{\omega^{\omega^{n+1}}}\}_{n<\omega}$ is the difference hierarchy over $\mathcal{L}_2$, and so on.  

The FH over the effective Borel base will be denoted by  $\{\Sigma_\alpha(X)\}_{\alpha<\varepsilon_0}$ and called the {\em effective Wadge hierarchy in $X$}. 
The corresponding boldface sequence $\{\mathbf{\Sigma}_\alpha(\mathcal{N})\}_{\alpha<\varepsilon_0}$  studied in \cite{s03}, forms a small but important fragment of the classical Wadge hierarchy in the Baire space. Some fragments of the classical Wadge hierarchy in quasi-Polish spaces were first defined and studied in \cite{s17} and recently extended  to all levels providing the boldface version of results of this paper for all levels of the Wadge hierarchy in quasi-Polish spaces \cite{s19}.

Some special cases of the EWH were considered before (see \cite{s08} and references therein). E.g., for the case $X=\mathbb{N}$ the  hierarchy $\{\Sigma_\alpha(\mathbb{N})\}_{\alpha<\varepsilon_0}$ was introduced  in \cite{s83}  by iterating of some jump operators on the many-one degrees; later an equivalent characterisation via set operations was obtained and used to classify many natural index sets. The FH over the base $(\mathcal{R}\cap\mathbf{\Sigma}^0_1(A^\omega), \mathcal{R}\cap\mathbf{\Sigma}^0_2(A^\omega),BC(\mathcal{R}\cap\mathbf{\Sigma}^0_2(A^\omega)),BC(\mathcal{R}\cap\mathbf{\Sigma}^0_2(A^\omega)),\ldots)$, where $\mathcal{R}$ is the class of regular omega-languages, coincides with the famous Wagner hierarchy which was introduced independently by K. Wagner in different terms.

\subsection{Computable quasi-Polish spaces}\label{cqp}

Though the effective hierarchies are naturally defined in arbitrary effective space, some important properties only hold for special classes of spaces, identification of which was itself a non-trivial task. Recall  a similar situation in classical DST where the  spaces with ``good'' DST (namely, the quasi-Polish spaces) were identified relatively recently \cite{br}.
Quasi-Polish spaces \cite{br} have several  characterisations.
Effectivizing one of them we obtain the following notion  identified implicitly in \cite{s15} and explicitly in \cite{br1,hs}.

\begin{Definition}\label{effqp}
By a  computable quasi-Polish space  we mean an effective space  $(X,\beta)$ such that there exists a computable effectively open surjection $\xi:\mathcal{N}\to X$ from the Baire space onto $(X,\beta)$.
 \end{Definition}

As shown in \cite{s15,br1,hs},  CQP-spaces do satisfy effective versions of several important properties of quasi-Polish spaces. E.g. they  subsume  computable Polish spaces and  computable domains and satisfy the effective Hausdorff and Suslin-Kleene theorems. The class of CQP-spaces includes most  of cb$_0$-spaces considered in the literature. In particular, all spaces mentioned in Section \ref{efclass} are CQP-spaces. 

In this paper we establish some good properties of the EWH in CQP-spaces. For this we use, in particular, the following corollary of Theorem 3.1 \cite{ch} (extending Lemma 3.1 in \cite{hs}) which effectivises the corresponding classical fact \cite{sr07,br} and shows that Baire category technique is consistent with effectivity. 
For any continuous function $f:X\to Y$ and $S\subseteq X$, let $f[S]$ consist of all $y\in Y$ such that $S\cap f^{-1}(y)$ is not meager in $f^{-1}(y)$. Please be careful in distinguishing $f[S]$ and the image $f(A)$.

\begin{Proposition}\cite{ch}\label{baire}
Let $f:X\to Y$ be a computable effectively open surjection between effective  spaces. Then $f[S]\in\Sigma^0_n(Y)$ for every $S\in\Sigma^0_n(X)$, and, for every $A\subseteq Y$, $f^{-1}(A)\in\Sigma^0_n(X)$ iff $A\in\Sigma^0_n(Y)$. 
\end{Proposition}

\section{Fine hierarchy of $k$-partitions}\label{fhkpart}

Here we discuss the FHs not only of subsets of $X$ but also  of $k$-partitions  $A:X\to\{0,\ldots,k-1\}=\bar{k}$ which may also be written as $k$-tuples $(A_0,\ldots,A_{k-1})$ where $A_i=A^{-1}(i),i<k$. To avoid trivialities, we always assume that $k\geq2$. Note that $2$-partitions of $X$ are essentially subsets of $X$. This section is mostly a reminder from \cite{s12}, with small modifications and additions.

Surprisingly, the extension from sets to $k$-partitions simplifies many definitions and  proofs related to the FH. The reason is that defining   FHs of sets by induction on ordinals leads to tedious inductive proofs while the FHs of $k$-partitions may be defined using intuitively clear manipulation with labeled trees; the corresponding proofs do not use induction on ordinals which arises only once, to establish the relation of labeled trees to ordinals for $k=2$  (see Section 8 of \cite{s12} for additional details). We first sketch this relation and the tree approach informally, and then recall the precise definitions.

Let $(Q;\leq)$ be a preorder. A {\em $Q$-tree}  is a pair $(T,t)$ consisting of a finite tree $T\subseteq\omega^*$  and a labeling $t:T\to Q$. Let  ${\mathcal T}_Q$ denote the set of all finite $Q$-trees.  The {\em $h$-preorder} $\leq_h$ on  ${\mathcal T}_Q$ is defined as follows: $(T,t)\leq_h (S,s)$, if there is a monotone function
$f:(T;\sqsubseteq)\to(S;\sqsubseteq)$ satisfying $\forall
x\in T(t(x))\leq s(f(x)))$. Though many results of this paper may be extended to arbitrary finite preorders $Q$ in place of $\bar{k}$, we will mainly stick to $k$-partitions in order to avoid some complications and exceptions.

The preorder $Q$ is a {\em well quasiorder} (WQO) if it has neither infinite descending chains nor infinite antichains. An example of WQO is the antichain $\bar{k}$ with $k$ elements. A famous Kruskall's theorem implies that if $Q$ is WQO then $({\mathcal T}_Q;\leq_h)$ is WQO. Define the sequence
$\{\mathcal{T}_k(n)\}_{n<\omega}$ of preorders by induction on $n$
as follows: $\mathcal{T}_k(0)=\overline{k}$  and
$\mathcal{T}_k(n+1)=\mathcal{T}_{\mathcal{T}_k(n)}$. The sets $\mathcal{T}_k(n)$, $n<\omega$, are pairwise disjoint but, identifying
the elements $i$ of $\overline{k}$ with the corresponding singleton trees $s(i)$   labeled by $i$ (which are precisely the
minimal elements  of $\mathcal{T}_k(1)$), we may think that
$\mathcal{T}_k(0)\sqsubseteq\mathcal{T}_k(1)$, i.e. the quotient-poset of the first preorder is an initial segment of the quotient-poset of the other. This also induces an embedding of $\mathcal{T}_k(n)$ into $\mathcal{T}_k(n+1)$ as an initial segment, so (abusing notation) we may think that $\mathcal{T}_k(0)\sqsubseteq\mathcal{T}_k(1)\sqsubseteq\cdots$, hence
$\mathcal{T}_k(\omega)=\bigcup_{n<\omega}\mathcal{T}_k(n)$ is WQO w.r.t. the induced preorder which we also denote $\leq_h$. The embedding $s$ is extended to $\mathcal{T}_k(\omega)$ by defining $s(T)$ as the singleton tree labeled by $T$. 

With any base $\mathcal{L}=\{\mathcal{L}_n\}_{n<\omega}$ in $X$ we then associate the {\em fine hierarchy of $k$-partitions over $\mathcal{L}$} which is a family $\{\mathcal{L}(X,T)\}_{T\in\mathcal{T}_k(\omega)}$ of subsets of $k^X$ introduced in \cite{s12} (we will recall the definition later in this section). As shown in \cite{s12}, $T\leq_hS$ implies $\mathcal{L}(X,T)\subseteq\mathcal{L}(X,S)$, hence $(\{\mathcal{L}(X,T)\mid T\in\mathcal{T}_k(\omega)\};\subseteq)$ is WQO. 

The FH of sets from Subsection \ref{efwadge} is obtained from this construction for $k=2$ since the quotient-poset of $(\mathcal{T}_2(\omega);\leq_h)$ has order type $\bar{2}\cdot\varepsilon_0$, i.e. for some $T_\alpha, T'_\alpha\in\mathcal{T}_2(\omega)$ we have: $T_\alpha, T'_\alpha<_hT_\beta, T'_\beta$ for $\alpha<\beta<\varepsilon_0$, and any element of $\mathcal{T}_2(\omega)$ is $h$-equivalent to precisely one of $T_\alpha, T'_\alpha$. We then set $\mathcal{S}_\alpha=\mathcal{L}(X,T_\alpha)$ for all $\alpha<\varepsilon_0$. For details see Definition 8.27 and Proposition 8.28 in \cite{s12}.

We come back to precise definitions. 
For any finite tree $T\subseteq\omega^*$ and any $T$-family $\{U_\tau\}$ of subsets of $X$, we define the $T$-family $\{\tilde{U}_\tau\}$ of subsets of $X$ by $\tilde{U}_\tau=U_\tau\setminus\bigcup\{U_{\tau'}\mid\tau\sqsubset\tau'\in T\}$. The $T$-family $\{U_\tau\}$ is {\em monotone} if $U_\tau\supseteq U_{\tau'}$ for all $\tau\sqsubseteq\tau'\in T$. We associate with any $T$-family $\{U_\tau\}$ the monotone $T$-family $\{U'_\tau\}$ by $U'_\tau=\bigcup_{\tau'\sqsupseteq\tau}U_{\tau'}$. A $T$-family $\{V_\tau\}$  is {\em reduced} if it is monotone and satisfies $V_{\tau i}\cap V_{\tau j}=\emptyset$ for all $\tau i,\tau j\in T$. Obviously, for any reduced $T$-family $\{V_\tau\}$  the components $\tilde{V}_\tau$ are pairwise disjoint. 

Recall that a class of sets $\mathcal{C}\subseteq P(X)$ has the {\em reduction property}  if for any $C_0,C_1\in\mathcal{C}$   there are disjoint $R_0,R_1\in\mathcal{C}$ such that $R_0\subseteq C_0,R_1\subseteq C_1$ and $R_0\cup R_1=C_0\cup C_1$. Note that if $\mathcal{C}\subseteq P(X)$ has the  reduction property then  for any finite sequence $C_0,\ldots,C_n\in\mathcal{C}$   there are pairwise disjoint $R_0\ldots,R_n\in\mathcal{C}$ such that $R_i\subseteq C_i$ for every $i\leq n$, and $R_0\cup\cdots\cup R_n=C_0\cup\cdots\cup  C_n$; we call such a sequence $(R_0\ldots,R_n)$  a {\em reduct} of $(C_0,\ldots,C_n)$.

The class $\mathcal{C}\subseteq P(X)$ has the {\em $\sigma$-reduction property} if for every infinite sequence $C_0,C_,\ldots$ in $\mathcal{C}$   there is an infinite sequence  $R_0,R_1\ldots$ in $\mathcal{C}$ with  pairwise disjoint elements such that $R_i\subseteq C_i$ for every $i$, and $\bigcup_iR_i=\bigcup_iC_i$; we call such a sequence $R_0,R_1\ldots$  a {\em $\sigma$-reduct} of $C_0,C_,\ldots$. Obviously, $\sigma$-reduction property implies the reduction property.

In Section \ref{trans} we will need an effective version of the $\sigma$-reduction property. This requires to fix a numbering $\gamma$ of $\mathcal{C}$ (i.e., a surjection $\gamma:\mathbb{N}\to\mathcal{C}$). We say that $\mathcal{C}\subseteq P(X)$ has the {\em effective $\sigma$-reduction property w.r.t. $\gamma$} if  every infinite sequence  in $\mathcal{C}$ computable in $\gamma$ (i.e., $C_i=\gamma_{f(i)}$ for some computable function $f$ on $\mathbb{N}$) has a $\sigma$-reduct which is also computable in $\gamma$; moreover, the $\sigma$-reduct should be found uniformly.

\begin{Lemma}\label{reduce}
\begin{enumerate}
\item Let  $\{U_\tau\}$ be a $T$-family of ${\mathcal L}_n$-sets. Then $\bigcup_\tau U_\tau=\bigcup_\tau\tilde{U}_\tau$, $\tilde{U}_\tau=\widetilde{U'}_\tau\in{\mathcal L}_{n+1}\cap\check{\mathcal L}_{n+1}$, and $\tilde{U}_\tau\cap\tilde{U}_{\tau'}=\emptyset$ for $\tau\sqsubset\tau'\in T$.
\item Let   $\{U_\tau\}$ be a monotone $T$-family of ${\mathcal L}_n$-sets, and ${\mathcal L}_n$ has the reduction property. Then there is a reduced $T$-family $\{V_\tau\}$ of ${\mathcal L}_n$-sets such that $V_\tau\subseteq U_\tau$ and $\bigcup_i\{V_{\tau i}\mid \tau i\in T\}=\bigcup_i\{U_{\tau i}\mid \tau i\in T\}$, and $\tilde{V}_\tau\subseteq \tilde{U}_\tau$ for each $\tau\in T$.
\end{enumerate}
\end{Lemma}

{\em Proof.} 1. The proof is straightforward. For instance, the set $\tilde{U}_\tau$ is, by definition, a difference of ${\mathcal L}_n$-sets, hence $\tilde{U}_\tau\in{\mathcal L}_{n+1}\cap\check{\mathcal L}_{n+1}$.

2. The property is checked by a top-down (assuming that trees grow downwards) application of the reduction property.  If $T=\{\varepsilon\}$ is singleton, there is nothing to prove. Otherwise, let $\{V_i\}$ be a reduct of $\{U_i\}$ and let $U'_{i\tau}=V_i\cap U_{i\tau}$ for all $i\tau\in T$. Apply this procedure to the trees $T(i)=\{\sigma\mid i\sigma\in T\}$ and further downwards whenever possible. Since $T$ is finite, we will finally obtain a desired reduced family which we call a reduct of $\{U_\tau\}$. It has the desired properties.
 \qed

With any base $\mathcal{L}(X)$  we  associate some other bases as follows. For any $m<\omega$, let $\mathcal{L}^m(X)=\{\mathcal{L}_{m+n}(X)\}_n$; we call this base  the {\em $m$-shift of $\mathcal{L}(X)$}.  For any $U\in\mathcal{L}_0$, let $\mathcal{L}(U)=\{\mathcal{L}_n(U)\}_{n<\omega}$ where $\mathcal{L}_n(U)=\{U\cap S\mid S\in\mathcal{L}_{n}(X)\}$; we call this base  the {\em $U$-restriction of $\mathcal{L}(X)$}.

We will use the following technical notions. The first one is the notion ``$F$ is a  $T$-family in $\mathcal{L}(X)$'' which is defined by induction as follows.

\begin{Definition}\label{famin}
\begin{enumerate}
\item If $T\in\mathcal{T}_k(0)$ then $F=\{X\}$.
\item If $(T,t)\in\mathcal{T}_k(n+1)$ then $F=(\{U_\tau\},\{F_\tau\})$ where $\{U_\tau\}$ is a monotone $T$-family of  ${\mathcal L}_0$-sets with $T_\varepsilon=X$ and, for each $\tau\in T$, $F_\tau$ is  a  $t(\tau)$-family in $\mathcal{L}^1(\tilde{U}_\tau)$.
\end{enumerate}
\end{Definition}

The notion of a reduced family is obtained from this definition by requiring $\{U_\tau\}$ and $F_\tau$ in item (2) to be reduced.

The second one is the notion ``a  $T$-family $F$ in $\mathcal{L}(X)$ determines a partition $A:X\to\bar{k}$'' which is defined by induction as follows. (In general, not every family determines a $k$-partition but every reduced family does.)

\begin{Definition}\label{determ}
\begin{enumerate}
\item If $T\in\mathcal{T}_k(0)$, $T=i<k$ (so $F=\{X\}$), then $T$ determines the constant partition $A=\lambda x.i$.
\item If $(T,t)\in\mathcal{T}_k(n+1)$ (so $F$ is of the form $(\{U_\tau\},\{F_\tau\})$) then $T$ determines the $k$-partition $A$ such that $A|_{\tilde{U}_\tau}=B_\tau$  for every $\tau\in T$, where $B_\tau:\tilde{U}_\tau\to\bar{k}$ is the $k$-partition of $\tilde{U}_\tau$ determined by $F_\tau$. 
\end{enumerate}
\end{Definition}

Since inductive proofs according the given definitions sometimes hide the ideas, let us give examples of explicit descriptions of the introduced notions. For $T=i\in\mathcal{T}_k(0)$, there is  only one $T$-family $\{X\}$ in   $\mathcal{L}(X)$ which determines the constant partition $\lambda x.i$. For $T\in\mathcal{T}_k(1)$, a $T$-family $F$ in   $\mathcal{L}(X)$ is essentially a monotone family $\{U_\tau\}$ of $\mathcal{L}_0(X)$-sets whose components $\tilde{U}_\tau$ cover $X$. Such a family determines $A$ if $A(x)=t(\tau)$, for any $\tau\in T$ with $x\in\tilde{U}_\tau$. Note that $t:T\to\bar{k}$ and that $x$ may belong to different components $\tilde{U}_\tau$, $\tilde{U}_\sigma$ with incomparable $\tau,\sigma$.

For $T\in\mathcal{T}_k(2)$, a $T$-family $F$ in   $\mathcal{L}(X)$ consists of a family $\{U_\tau\}$ as above, and, for each $\tau_0\in T$, a family $\{U_{\tau_0\tau_1}\}_{\tau_1\in t_0(\tau_0)}$ of $\mathcal{L}_1(X)$-sets whose components (which we call second-level components) $\tilde{U}_{\tau_0\tau_1}$ cover $\tilde{U}_{\tau_0}$  (called first-level components). Such an $F$ determines $A$ if $A(x)=t_1(\tau_1)$, for all $\tau_0\in T,\tau_1\in t_0(\tau_0)$ with $x\in\tilde{U}_{\tau_0\tau_1}$. Note that $t_0:T\to\mathcal{T}_k(1),t_1:t_0(\tau_0)\to\bar{k}$.

For $T\in\mathcal{T}_k(3)$, a $T$-family $F$ in   $\mathcal{L}(X)$ consists of  families $\{U_\tau\},\{U_{\tau_0\tau_1}\}$ as above and, for all $\tau_0\in T,\tau_1\in t_0(\tau_0)$, a family $\{U_{\tau_0\tau_1\tau_2}\}_{\tau_2\in t_1(\tau_1)}$ of $\mathcal{L}_2(X)$-sets whose components $\tilde{U}_{\tau_0\tau_1\tau_2}$ of the third level cover $\tilde{U}_{\tau_0\tau_1}$. Such $F$ determines $A$ if $A(x)=t_2(\tau_2)$, for all $\tau_0\in T,\tau_1\in t_0(\tau_0),\tau_2\in t_1(\tau_1)$ with $x\in\tilde{U}_{\tau_0\tau_1\tau_2}$. Note that $t_0:T\to\mathcal{T}_k(2),t_1:t_0(\tau_0)\to\mathcal{T}_k(1),t_2:t_1(\tau_1)\to\bar{k}$.

Intuitively, the $T$-family $F$ (say, in an effective Borel base) that determines $A$ provides a mind-change algorithm for computing $A(x)$ for a given $x\in X$ as follows. First, we search for a component $\tilde{U}_{\tau_0}$ containing $x$; this is the usual mind-change procedure working with differences of $\Sigma^0_1$-sets. While $x$ sits in $\tilde{U}_{\tau_0}$, we search  for a component $\tilde{U}_{\tau_0\tau_1}$ containing $x$; this is a harder  mind-change procedure working with differences of $\Sigma^0_2$-sets, and so on. Note that if $F$ is reduced then the computation is ``linear'' since the components of each level are pairwise disjoint, otherwise the algorithm is ``parallel'' since already at the first level $x$ may belong to several components $\tilde{U}_{\tau_0}$. With this interpretation in mind, we consider the (effective) Wadge hierarchy as an ``iterated difference hierarchy'', one should only make precise how to ``iterate'' them. 

We continue this process of computing $A(x)$ until we reach a sequence  $(\tau_0,\ldots,\tau_m)$ such that $x\in\tilde{U}_{\tau_0\cdots\tau_m}$ and $t_m(\tau_m)<k$; such sequences  $(\tau_0,\ldots,\tau_m)$ and components $\tilde{U}_{\tau_0\cdots\tau_m}$ are called {\em terminating}. Note that the terminating components cover $X$ and that $A^{-1}(i)$ is the the union of all terminating components $\tilde{U}_{\tau_0\cdots\tau_m}$ with $t_m(\tau_m)=i$, for every $i<k$.

Note that if $F$ is reduced then all families of sets $\{\tilde{U}_{\tau_0}\},
\{\tilde{U}_{\tau_0\tau_1}\}_{\tau_1},
\{\tilde{U}_{\tau_0\tau_1\tau_2}\}_{\tau_2},\ldots$ are pairwise disjoint,  all components of any fixed level form a partition of $X$, and all terminating components form a partition of $X$.

Let $F=(\{U_{\tau_0}\},\{U_{\tau_0\tau_1}\},\ldots)$ and $G=(\{V_{\tau_0}\},\{V_{\tau_0\tau_1}\},\ldots)$ be  $T$-families in $\mathcal{L}(X)$ unravelled as discussed above. We say that $G$ {\em is a reduct of $F$} if $G$ is  reduced and $\tilde{V}_{\tau_0\cdots\tau_m}\subseteq\tilde{U}_{\tau_0\cdots\tau_m}$ for every terminating $(\tau_0,\ldots,\tau_m)$. 

\begin{Lemma}\label{reduct}
Let ${\mathcal L}(X)$ be a reducible base in $X$ and let $T\in\mathcal{T}_k(\omega)$. Then any $T$-family $F$ in ${\mathcal L}(X)$ has a reduct $G$. Moreover, if $F$ determines $A$ then any reduct of $F$ also determines $A$.
\end{Lemma}

{\em Proof Sketch.} We follow the procedure of computing $A(x)$ described above. If  $T$ is a singleton tree, we set $G=F=\{X\}$; then $F,G$ determine the same constant $k$-partition.  Otherwise, $F$ has the form as above. Let  $G$  be obtained from $F$ by repeated reductions from Lemma \ref{reduce}(2), so in particular $\tilde{V}_{\tau_0\cdots\tau_m}\subseteq\tilde{U}_{\tau_0\cdots\tau_m}$ for each terminating  $(\tau_0,\ldots,\tau_m)$.

For the second assertion, let $F$ determine $A$ and let $G$ be a reduct of $F$. For any $x\in X$, let $\tilde{V}_{\tau_0\cdots\tau_m}$ be the unique terminating component of $G$ containing $x$. Then also $x\in\tilde{U}_{\tau_0\cdots\tau_m}$, hence $A(x)=t_m(\tau_m)$ and $G$ determines $A$.
 \qed
 
\begin{Lemma}\label{tfaml}
 Every $T$-family in ${\mathcal L}(X)$ determines at most one $k$-partition of $X$.
 Every reduced $T$-family in ${\mathcal L}(X)$ determines precisely one $k$-partition of $X$.
\end{Lemma}

{\em Proof.} The second assertion follows from the remark that the terminating components of $G$ form a partition of $X$. For the first assertion, let a $T$-family $F$ in ${\mathcal L}(X)$ determine $k$-partitions $A,B$ of $X$ and let $x\in X$. If  $T$  is a singleton tree, $F$ determines a constant $k$-partition, so in particular $A(x)=B(x)$.  Otherwise, $F=(\{U_\tau\},\{F_\tau\})$ as specified above. By the procedure of computing $A(x)$, there is a terminating component $\tilde{U}_{\tau_0\cdots\tau_m}\ni x$ of $F$. By Definition \ref{determ}, $A(x)=t_m(\tau_m)=B(x)$.
 \qed

We are ready to give a precise definition of  the FH of $k$-partitions over ${\mathcal L}(X)$.

\begin{Definition}\label{fh}
The fine hierarchy of $k$-partitions over ${\mathcal L}(X)$ is the family $\{\mathcal{L}(X,T)\}_{T\in\mathcal{T}_k(\omega)}$ of subsets of $k^X$ where ${\mathcal L}(X,T)$ is the set of $A:X\to\bar{k}$ determined by some $T$-family in $\mathcal{L}(X)$. 
\end{Definition}

As already mentioned above, $T\leq_hS$ implies $\mathcal{L}(X,T)\subseteq\mathcal{L}(X,S)$, hence $(\{\mathcal{L}(X,T)\mid T\in\mathcal{T}_k(\omega)\};\subseteq)$ is WQO. See   \cite{s12} for additional technical details.

\section{Effective  Wadge hierarchy of $k$-partitions}\label{wkpart}

Here we discuss properties of the EWH in an effective space, i.e. the FH  over the effective Borel bases $\mathcal{L}(X)=\{\Sigma^0_{1+n}(X)\}$. To stress that it is a very special case of the FH we denote it in a special way: $\{\Sigma(X,T)\}_{T\in\mathcal{T}_k(\omega)}$ instead of the general notation $\{\mathcal{L}(X,T)\}_{T\in\mathcal{T}_k(\omega)}$.  

From now on, all the discussed bases are the effective Borel bases only. They have some specific features, e.g. any level $\Sigma^0_{1+n}(X)$ is closed under effective countable unions. 
We formulate some more properties. A base $\mathcal{L}(X)$ is {\em reducible (resp. $\sigma$-reducible, effectively $\sigma$-reducible)} if every its level has the corresponding property. By a {\em morphism} $g:\mathcal{L}(X)\to\mathcal{L}(Y)$ between effective Borel bases we mean a function $g:P(X)\to P(Y)$ such that any restriction $g|_{\Sigma^0_{1+n}(X)}$ is a computable function from $\Sigma^0_{1+n}(X)$ to $\Sigma^0_{1+n}(Y)$ which preserves effective countable unions and satisfies $g(\emptyset)=\emptyset$, $g(X)=Y$. Obviously, the identity function on $P(X)$ is a morphism of $\mathcal{L}(X)$ to itself, and if $g:\mathcal{L}(X)\to\mathcal{L}(Y)$ and $h:\mathcal{L}(Y)\to\mathcal{L}(Z)$ are morphisms  then $h\circ g:\mathcal{L}(X)\to\mathcal{L}(Z)$ is also a morphism.

We associate with any $T$-family $F$ in $\mathcal{L}(Y)$ the $T$-family $f^{-1}(F)$   in $\mathcal{L}(X)$ by induction as follows: if $T\in\mathcal{T}_k(0)$ and $F=\{Y\}$, then set $f^{-1}(F)=\{X\}$;
 if $(T,t)\in\mathcal{T}_k(n+1)$ and $F=(\{U_\tau\},\{F_\tau\})$) then set $f^{-1}(F)=(\{f^{-1}(U_\tau)\},\{f^{-1}(F_\tau)\})$.

\begin{Proposition}\label{prop}
\begin{enumerate}
\item The 1-shift  of every effective Borel base  is effectively $\sigma$-reducible.
\item For any $X\in\{\mathbb{N},\mathcal{N},A^\omega\}$, the  effective Borel base $\mathcal{L}(X)$ is effectively $\sigma$-reducible.
\item Let $f:X\to Y$ be a computable function between effective spaces. Then $f^{-1}:\mathcal{L}(Y)\to \mathcal{L}(X)$ is a morphism of effective Borel bases, and $A\in\Sigma(Y,T)$ implies $A\circ f\in\Sigma(X,T)$, for every $T\in\mathcal{T}_k(\omega)$.
\end{enumerate}
\end{Proposition}

{\em Proof.} (1) It is well known (see e.g. \cite{ke95,s13}) that any level $\bfSig^0_{2+\alpha}(X)$, $\alpha<\omega_1$, of the Borel hierarchy in a cb$_0$-space $X$ has the $\sigma$-reduction property. A straightforward effectivisation of the proof shows that any level $\Sigma^0_{2+n}(X)$ of the effective Borel hierarchy in an effective space $X$ has the effective  $\sigma$-reduction property which implies the assertion.

(2) It is well known (see e.g. \cite{ke95,s13}) that if $X$ is a zero-dimensional cb$_0$-space then $\bfSig^0_1(X)$ has the $\sigma$-reduction property. A straightforward effectivisation of the proof shows that  $\Sigma^0_{1}(X)$ has the effective  $\sigma$-reduction property  for each $X\in\{\mathbb{N},\mathcal{N},A^\omega\}$, which together with item (1) implies the assertion.

(3) Both assertions are straightforward,  let us consider e.g.  the second one. Let $A\in\Sigma(Y,T)$, then $A$ is determined by a $T$-family $F$ in $\mathcal{L}(Y)$. Clearly, $A\circ f$ is then determined by the $T$-family $f^{-1}(F)$ in $\mathcal{L}(X)$. Therefore, $A\circ f\in\Sigma(X,T)$.
\qed

The reduction property is  crucial for understanding which levels of the EWHs have ``good'' numberings and complete sets (or $k$-partitions). By the {\em effective Wadge reducibility} in an effective space $X$ we mean the many-one reducibility $\leq^X_W$ by computable functions on $X$; this reducibility applies not only to subsets of $X$ but also to any functions defined on $X$ (in particular, to $k$-partitions). 

To define the ``good'' numberings, we need effective versions of some notions from \cite{s13,s17}. By {\em effective family of pointclasses} we mean a family $\{\Gamma(X)\}_X$ parametrised by  the effective spaces $X$ such that $\Gamma(X)\subseteq P(X)$, and $f^{-1}:\Gamma(Y)\to\Gamma(X)$ for any computable $f:X\to Y$. A numbering $\nu:\mathbb{N}\to\Gamma(X)$ is {\em $\Gamma$-computable} if $\{(n,x)\mid x\in\nu(n)\}\in\Gamma(\mathbb{N}\times X)$. Such a numbering is {\em principal} if any $\Gamma$-computable numbering $\mu$ is reducible to $\nu$, i.e. $\mu=\nu\circ f$ for a computable function $f$ on $X$. 

The corresponding notions for $k$-partitions are defined in a similar way. Namely, a {\em family of effective $k$-partition classes} is a family $\{\Gamma(X)\}_X$ parametrised by the effective spaces such that $\Gamma(X)\subseteq k^X$ for each $X$, and $A\circ f\in\Gamma(X)$ for every computable function $f:X\to Y$ and every $A\in\Gamma(Y)$. A numbering $\nu:\mathbb{N}\to\Gamma(X)$ is a {\em $\Gamma$-numbering} if its
{\em universal $k$-partition} $(a,x)\mapsto\nu(a)(x)$ is in $\Gamma(\mathbb{N}\times X)$, and $\nu$ is a {\em principal $\Gamma$-numbering} if it is a $\Gamma$-numbering and any $\Gamma$-numbering  $\mu:\mathbb{N}\to\Gamma(X)$ is reducible to $\nu$.  Note that if $\nu:\mathbb{N}\to\Gamma(X)$ is principal then it is a surjection and that $\Gamma(X)$ has at most one principal numbering, up to the equivalence of numberings (recall that two numberings are equivalent if they are reducible to each other).

It is easy to check that all $\Sigma$-levels of the effective hierarchies in Subsection \ref{efhier} form effective pointclasses, and their standard numberings are principal computable. The next result partially extends this to the EWH.

\begin{Proposition}\label{prop1}
\begin{enumerate}
\item For any $T\in\mathcal{T}_k(\omega)$ and any effective space $X$, the  level $\Sigma(X,s(T))$ has a principal computable numbering. In particular, this holds  for any level $\Sigma_{\omega^\alpha(X)}$ of the effective Wadge hierarchy of sets.
\item If $X\in\{\mathbb{N},\mathcal{N}\}$ then the assertion (1) holds for all levels of the effective Wadge hierarchies, and any level has a  $\leq^X_W$-complete set.
\end{enumerate}
\end{Proposition}

{\em Proof Sketch.} Consider e.g. the level  $\Sigma(\mathcal{N},T)$  in (2), the other cases are similar. Let $F_0,F_1,\ldots$ be the computable numbering of $T$-families in $\mathcal{L}(\mathcal{N})$ induced by the numberings of levels of the effective Borel hierarchy. For any $i$, let $F_i'$ be a reduced family obtained from $F_i$ by the procedure of top-down application of the reduction property. Observe that $F_i'=F_i$ if $F_i$ was already a reduced family, and that, by Lemma \ref{reduce}, if $F_i$ determines $A$ then so does $F_i'$. By Lemma \ref{reduce}(2) and Proposition \ref{prop}, $F_0',F_1',\ldots$ is a computable numbering of the reduced $T$-families  in $\mathcal{L}(\mathcal{N})$. Then $A_0,A_1,\ldots$, where $A_i$ is determined by $F_i'$, is a desired numbering of $\Sigma(\mathcal{N},T)$. The $\Sigma(\mathcal{N},T)$-complete $k$-partition is  obtained by essentially taking the disjoint union of $A_0,A_1,\ldots$ (using a computable homeomorphism between $\mathbb{N}\times\mathcal{N}$ and $\mathcal{N}$).
 \qed

An important question about any hierarchy of sets is the question on whether it collapses or not. For instance, for an EWH of sets $\{\Sigma_\alpha(X)\}_{\alpha<\varepsilon_0}$ we say that it {\em does not collapse} if $\Sigma_\alpha(X)\not\subseteq\Pi_\alpha(X)$ for all $\alpha<\varepsilon_0$. This definition readily extends to the EWH $\{\Sigma(X,T)\}_{T\in\mathcal{T}_k(\omega)}$ of $k$-partitions: we say that this hierarchy {\em does not collapse} if $\Sigma(X,T)\not\subseteq\Sigma(X,S)$ for all $T,S\in\mathcal{T}_k(\omega)$ with $T\not\leq_hS$. This is equivalent to saying that the partial order $(\{\Sigma(X,T)\mid T\in\mathcal{T}_k(\omega)\};\subseteq)$ is isomorphic to $\bar{2}\cdot\varepsilon_0$.

From the results in \cite{s83} (and the equivalence of hierarchy there to the FH in $\mathbb{N}$) it follows that the EWH of $k$-partitions in $\mathbb{N}$ does not collapse. It might also be shown that the EWH of $k$-partitions in $X$ does not collapse for every $X\in\{P\omega,\mathcal{N},A^\omega\}$. Though we do not plan in this paper to study the collapse problem systematically, we note that the preservation property considered in the next section has an interesting application to this problem.

\section{Preservation of levels}\label{preserve}

A main result of this paper is the following preservation property for levels of the EWH.

\begin{Theorem}\label{main}
Let $f:X\to Y$ be a computable effectively open surjection between  effective spaces and $A: Y\to\bar{k}$. Then for any $T\in\mathcal{T}_k(\omega)$ we have: $A\in\Sigma(Y,T)$ iff $A\circ f\in\Sigma(X,T)$. In particular, for all $A\subseteq Y$ and $\alpha<\varepsilon_0$ we have: $A\in\Sigma_\alpha(Y)$ iff $f^{-1}(A)\in\Sigma_\alpha(X)$.
\end{Theorem}

We fix $f$ as in the formulation above and first  prove two lemmas about the function $A\mapsto f[A]$ from Section \ref{cqp}
which  was used e.g. in \cite{sr07,br,s17}. 
   The first lemma follows  from Definitions \ref{famin},  \ref{determ} and Proposition \ref{baire}. Note that item (2) of this lemma is formulated for more general trees than the finite trees sufficient for this section; the more general form will be used in the next section.  

\begin{Lemma}\label{catim}
\begin{enumerate}
\item The function  $A\mapsto f[A]$ is a morphism from $\mathcal{L}(X)$ to $\mathcal{L}(Y)$, and $f[A]\subseteq f(A)$ for each $A\subseteq X$.
\item If $T$ is a c.e. well founded tree and $\{U_\tau\}$ is an effective $T$-family of $\Sigma^0_{1+n}(X)$-sets then $\{f[U_\tau]\}$ is an effective $T$-family of $\Sigma^0_{1+n}(Y)$-sets, and $\widetilde{f[U_\tau]}\subseteq f[\tilde{U}_\tau]$ for each $\tau\in T$.
\end{enumerate}
 \end{Lemma}

{\em Proof.} (1) The proof follows  from Proposition \ref{baire}.

(2) The first assertion follows from item (1), so we check the second one. Let $y\in\widetilde{f[U_\tau]}$, i.e. $y\in f[U_\tau]\setminus\bigcup\{f[U_{\tau'}]\mid \tau\sqsubset\tau'\in T\}$. Then $U_\tau\cap f^{-1}(y)$ is non-meager in $f^{-1}(y)$ and, for each $\tau\sqsubset\tau'\in T$, $U_{\tau'}\cap f^{-1}(y)$ is meager in $f^{-1}(y)$. Then $(\bigcup\{U_{\tau'}\mid \tau\sqsubset\tau'\in T\})\cap f^{-1}(y)$ is meager in $f^{-1}(y)$, hence $\tilde{U}_\tau=U_\tau\setminus\bigcup\{U_{\tau'}\mid \tau\sqsubset\tau'\in T\}$ is non-meager in $f^{-1}(y)$, i.e. $y\in f[\tilde{U}_\tau]$.
 \qed
 
We associate with any $T$-family $F$ in $\mathcal{L}(X)$ the $T$-family $f[F]$ in $\mathcal{L}(Y)$ by induction as follows: if $T\in\mathcal{T}_k(0)$ (hence $F=\{X\}$)  then we set $f[F]=\{Y\}$; if $T\in\mathcal{T}_k(n+1)$ (hence $F=(\{U_\tau\}_{\tau\in T},\{F_\tau\})$ and $t(\tau)\in\mathcal{T}_k(n)$ for each $\tau\in T$) then we set $f[F]=(\{f[U_\tau]\},\{f[F_\tau]\})$. That  $f[F]$ is really a $T$-family in $\mathcal{L}(Y)$, follows from Proposition \ref{baire} and Lemma \ref{catim}.

\begin{Lemma}\label{catim1}
Let $A:Y\to\bar{k}$ and let $A\circ f$ be determined by a $T$-family $F$ in $\mathcal{L}(X)$. Then $A$ is determined by the $T$-family $f[F]$. 
 \end{Lemma}
 
{\em Proof.} To simplify notation and to illustrate the ideas by typical example, we give a proof only for  $T\in\mathcal{T}_k(3)$ (see examples before Lemma \ref{tfaml}). Since $A\circ f$ is determined by $F$, $(A\circ f)(x)=t_2(\tau_2)$, for all $\tau_0\in T,\tau_1\in t_0(\tau_0),\tau_2\in t_1(\tau_1)$ with $x\in\tilde{U}_{\tau_0\tau_1\tau_2}$.

We have to show that $A$ is determined by $f[F]$, i.e. 
$A(y)=t_2(\tau_2)$, for all $y\in Y$ and $\tau_0\in T,\tau_1\in t_0(\tau_0),\tau_2\in t_1(\tau_1)$ with $y\in\widetilde{f[U_{\tau_0\tau_1\tau_2}]}$. 

For any given $y\in Y$, there exist $\tau_0,\tau_1,\tau_2$ with $y\in\widetilde{f[U_{\tau_0\tau_1\tau_2}]}$. By Lemma \ref{catim}(3), these conditions imply $y\in f[\tilde{U}_{\tau_0\tau_1\tau_2}]$, so $y=f(x)$ for some $x\in\tilde{U}_{\tau_0\tau_1\tau_2}$. Thus, $A(y)=(A\circ f)(x)=t_2(\tau_2)$.
 \qed

{\em Proof of Theorem \ref{main}.} Let $A\in\Sigma(Y,T)$. By Proposition \ref{prop}(3), $A\circ f\in\Sigma(X,T)$. Conversely, let $A\circ f\in\Sigma(X,T)$, then $A\circ f$ is determined by a $T$-family $F$ in $\mathcal{L}(X)$. By Lemma \ref{catim1}, $A$ is determined by the $T$-family $f[F]$ in $\mathcal{L}(Y)$, hence  $A\in\Sigma(Y,T)$.
 \qed 

The preservation theorem has several interesting corollaries. For instance, we have the following result on the non-collapse problem for EWH discussed at the end of Section \ref{wkpart}.
We define the preorder $\leq_{ceo}$ on effective spaces by: $Y\leq_{ceo}X$ if there is a computable effectively open surjection $f:X\to Y$. Obviously, $X$ is a CQP-space iff $X\leq_{ceo}\mathcal{N}$.

\begin{Proposition}\label{noncol}
If the EWH of $k$-partitions in an effective space $Y$ does not collapse and $Y\leq_{ceo}X$ then the EWH of $k$-partitions  in $X$ does not collapse.
\end{Proposition}

{\em Proof.} Let $\Sigma(Y,T)\not\subseteq\Sigma(Y,S)$ for all $T,S\in\mathcal{T}_k(\omega)$ with $T\not\leq_hS$ and let $f:X\to Y$ be a computable effectively open surjection. We have to show that $\Sigma(X,T)\not\subseteq\Sigma(X,S)$ for all $T,S\in\mathcal{T}_k(\omega)$ with $T\not\leq_hS$. Let $T\not\leq_hS$ and $A\in\Sigma(Y,T)\setminus\Sigma(Y,S)$. By Theorem \ref{main}, $f^{-1}(A)\in\Sigma(X,T)\setminus\Sigma(X,S)$.
 \qed

According to the  end of Section \ref{wkpart}, the EWH of $k$-partitions in  $\mathbb{N}$ does not collapse. Since $\mathbb{N}\leq_{ceo}\mathcal{N}$, we immediately obtain the following.

\begin{Corollary}\label{noncol1}
The EWH of $k$-partitions in the Baire space does not collapse.
\end{Corollary}

\section{Transfinite extensions}\label{trans}

Here we briefly discuss transfinite extensions of the effective hierarchies.
The transfinite extension of $\{\Sigma^0_n(X)\}_{n<\omega}$ is 
defined in a natural way (see e.g. \cite{s06}) as in  classical DST, only in place of $\omega_1$ 
one has to take the first non-computable ordinal $\omega_1^{CK}$. In
fact, to obtain reasonable effectivity properties one should
denote levels $\Sigma^0_{(a)}$ of the transfinite hierarchy not
by computable ordinals $\alpha<\omega_1^{CK}$ but rather by their
names $a,|a|_O=\alpha,$ in the Kleene notation system $(O;<_O)$
($a\mapsto|a|_O$ is a surjection from $O\subseteq\omega$ onto
$\omega_1^{CK}$), see chapter 16 of \cite{ro67}. The resulting {\em transfinite effective Borel hierarchy
$\{\Sigma^0_{(a)}(X)\}_{a\in O}$} is
extensional,  i.e. $\Sigma^0_{(a)}(X)=\Sigma^0_{(b)}(X)$ whenever
$|a|_O=|b|_O$.
The {\em transfinite effective Hausdorff hierarchy
$\{\Sigma^{-1,n}_{(a)}(X)\}_{a\in O}$ over}
$\Sigma^0_n(X)$ is also defined in a natural way  \cite{s06}). For $n=1$, we
abbreviate $\Sigma^{-1,1}_{(a)}(X)$ to $\Sigma^{-1}_{(a)}(X)$.
The effective Hausdorff hierarchy is not extensional.

These transfinite hierarchies were used to prove  effective versions of classical results from DST for arbitrary CQP-space $X$: the effective Suslin theorem  $\bigcup\{\Sigma^0_{(a)}(X)\mid
a\in O\}=\Delta^1_1(X)$ (follows from Theorem 4 in \cite{kk17}, see also Theorem 4 in \cite{s15}) and the effective Hausdorff theorem $\Delta^0_2(X)=\bigcup\{\Sigma^{-1}_{(a)}(X)\mid a\in O\}$ (Theorem 5 in \cite{s15}). This provides wide generalizations of the corresponding classical facts of computability theory: the Suslin-Kleene theorem is the first result for $X\in\{\mathbb{N},\mathcal{N}\}$ and the Ershov theorem is the second result for $X=\mathbb{N}$. Cf. the more general approach (based on represented sets) to transfinite extension and related results of DST in \cite{pa17}.

Since the EWH was defined above using  iterated labeled trees, it is natural to define transfinite extensions of this hierarchy  also in terms of trees. The natural choice is to consider  computable well founded trees in place of finite trees. To slightly simplify the proof of Theorem \ref{hausp} in the next section, we consider c.e. trees instead of computable trees, but the results  also hold for computable trees. 

Let $\mathcal{T}^*_k(\omega)$ be defined just as $\mathcal{T}_k(\omega)$ but using  c.e. well founded trees (and computable labeling functions) in place of finite trees. Definition \ref{famin} also makes sense under this change (only we have always require that the corresponding $T$-families of $\Sigma^0_{1+n}$-sets are uniform w.r.t. the principal computable numbering of $\Sigma^0_{1+n}$). In this way we also obtain the definition of levels $\Sigma(X,T)$ for arbitrary $\mathcal{T}^*_k(\omega)$. 
Repeating the proof of Theorem \ref{main} we obtain the following.

\begin{Proposition}\label{inftree}
Theorem \ref{main} remains true for $T\in\mathcal{T}^*_k(\omega)$. 
\end{Proposition} 

Due to the well known relation between computable and c.e. well founded trees to $\omega^{CK}_1$ and the Kleene notation system, the family $\{\Sigma(X,T)\}_{T\in\mathcal{T}^*_k(1)}$ for $k=2$ is essentially the same object as $\{\Sigma^{-1}_{(a)}(X)\}_{a\in O}$. For $k>2$, the family $\{\Sigma(X,T)\}_{T\in\mathcal{T}^*_k(1)}$ is a natural transfinite extension of the effective Hausdorff hierarchy of sets to $k$-partitions.

\section{Effective Hausdorff-Kuratowski theorem}\label{efHau}

Here we discuss effective versions of the  Hausdorff-Kuratowski (HK) theorem for $k$-partitions. We say that an effective space $X$ {\em satisfies $n$-HK theorem} if $\Delta^0_{n+2}(k^X)=\bigcup\{\Sigma(X,s^n(T))\mid T\in{\mathcal{T}^*_k(1)}\}$ where $\Delta^0_{n+2}(k^X)$ is the set of $A\in k^X$ with $A_0,\ldots,A_{k-1}\in\Delta^0_{n+2}(X)$ and $s^n$ is  the $n$th iteration of the function $s$  forming the singleton trees. For $n=0$ the equality simplifies to $\Delta^0_{2}(k^X)=\bigcup\{\Sigma(X,T)\mid T\in\mathcal{T}^*_k(1)\}$ which we call {\em the effective Hausdorff theorem for $k$-partitions}. A simple calculation shows that $n$-HK theorem is equivalent to $\Delta^0_{n+2}(k^X)\subseteq\bigcup\{\Sigma(X,s^n(T))\mid T\in{\mathcal{T}^*_k(1)}\}$ because the opposite inclusion clearly holds in every effective space.

\begin{Theorem}
If an effective space $X$ satisfies $n$-HK theorem  and $Y\leq_{ceo}X$ then so does $Y$. Thus, if $\mathcal{N}$ satisfies $n$-HK theorem  then so does every CQP-space.
\end{Theorem}

{\em Proof.} Let $f:X\to Y$ be a  computable effectively open surjection and $A\in\Delta^0_{n+2}(k^Y)$, $A=(A_0,\ldots,A_{k-1})$. Proposition \ref{prop}(3) implies that $A\circ f=(f^{-1}(A_0),\ldots,f^{-1}(A_{k-1}))\in\Delta^0_{n+2}(k^X)$, hence $A\circ f\in\Sigma(X,s^n(T))$ for some $T\in\mathcal{T}^*_k(1)$. By Proposition \ref{inftree}, $A\in\Sigma(Y,s^n(T))$.
 \qed
 
Thus, to prove the effective $n$-HK theorem for all CQP-spaces it suffices to prove it for the Baire space. Though we do not currently have a proof for $n>0$ we do have one for $n=0$. The next result extends Theorem 5 in \cite{s15} to $k$-partitions. 

\begin{Theorem}\label{hausp}
Every CQP-space satisfies the effective Hausdorff theorem for $k$-partitions.
\end{Theorem}  

{\em Proof.} We have to show that $\Delta^0_{2}(k^\mathcal{N})\subseteq\bigcup\{\Sigma(\mathcal{N},T)\mid T\in{\mathcal{T}^*_k(1)}\}$. Let $A\in\Delta^0_{2}(k^\mathcal{N})$, then $A$ is limit-computable, i.e. for some computable function $\Phi:\mathcal{N}\times\mathbb{N}\to\bar{k}$ we have $A(x)=\lim_n\Phi(x,n)$ (see e.g. Proposition 5.1 in \cite{s03} for $k=2$; the proof works for every $k$).

Let $M$ be an oracle Turing machine with $\Phi(x,n)=M^x(n)$. Define a uniformly c.e. sequence $R_0,R_1,\ldots$ of subsets of $\omega^*$ as follows. Let $R_0$ consist of the $\sqsubseteq$-minimal strings $\sigma$ such that the computation $M^\sigma(0)$ stops within $|\sigma|$ steps. Note that $R_0$ is a non-empty computable set whose elements are pairwise $\sqsubseteq$-incomparable. With any $\sigma\in R_0$ we associate the number $i_\sigma=0$. 
Suppose by induction that we already have $R_n$ and with any $\sigma\in R_n$ some $i_\sigma$ is associated such that $M^\sigma(i_\sigma)$ stops within $|\sigma|$ steps. Let $R_{n+1}$ consist of the $\sqsubseteq$-minimal strings $\tau$ such that $\sigma\sqsubset\tau$ for some $\sigma\in R_n$,  for some $i_\sigma< i<\tau$ the computation $M^\tau(i)$ stops within $|\tau|$ steps, and $M^\tau(i)\not=M^\sigma(i_\sigma)$. Let $i_\tau$ be the smallest such $i$. Note that $R_{n+1}$ is c.e. and already $R_1$ might be empty.

Let $T$ be the tree generated by $\bigcup_nR_n$, then $T$ is c.e. It is well founded because otherwise we would have an infinite sequence $\sigma_0\sqsubset\sigma_1\sqsubset\cdots$ such that $\sigma_i\in T_i$ for all $i$, hence for $x=sup\{\sigma_0,\sigma_1,\ldots\}$ the sequence $\{\Phi(x,n)\}_n$ changes infinitely often; a contradiction. We define the labeling $t:T\to\bar{k}$ as follows: if there is no $\sigma\in R_0$ with $\sigma\sqsubseteq\tau$ then $t(\tau)=0$, otherwise  $t(\tau)=M^\sigma(i_\sigma)$ where $\sigma\sqsubseteq\tau$ and $\sigma\in R_n$ for the largest possible $n$. The function $t$ is computable.
Define also the $T$-family $\{U_\tau\}$ of $\Sigma^0_1(\mathcal{N})$-sets by $U_\tau=[\tau]$. Then  this family determines $A$, hence $A\in\Sigma(\mathcal{N},T)$.
\qed

\section{Future work}\label{con}

To simplify notation, we concentrated in this paper on the finitary version of the EWH. Transfinite versions based on objects like the iterated labeled c.e. well founded trees seem adequate to develop transfinite versions of the EWH. We expect interesting extensions of the effective Hausdorff-Kuratowski theorem along these lines. Unfortunately, transfinite versions of EWH lead to cumbersome notation (see \cite{s83} where transfinite versions of EWH in $\mathbb{N}$ were considered).

If we consider arbitrary well founded labeled trees and their suitable iterations we obtain a broad extension of the classical Wadge hierarchy (see \cite{km19} and references therein). The methods of classical Wadge theory (including those in \cite{km19}) work only for zero-dimensional spaces. In \cite{s17} we suggested an approach to define and develop the Wadge hierarchy in arbitrary cb$_0$-spaces and demonstrated them to work for some initial segments of the Wadge hierarchy. Using methods of the present paper, we recently extended these partial results to the whole Wadge hierarchy, including the hierarchy of $k$-partitions. In particular, classical analogues of the results of this paper hold for arbitrary quasi-Polish spaces \cite{s19}.

A widely open question is to extend the methods and results of this paper beyond the effective Borel sets and $k$-partitions.

{\bf Acknowledgement.} This work was started in September 2019 during my visit to INRIA Nancy. I am grateful to M. Hoyrup and T. Kihara for useful discussions, and also to M. Hoyrup and INRIA for support and excellent research environment.


\begin{thebibliography}{aasa67}

%\bibitem{aj}   Abramsky S.,   Jung, A.: Domain theory. In: {\em Handbook of Logic in Computer Science}, v. 3, Oxford, 1994, 1--168.

\bibitem{ch}  A. Callard, M.  Hoyrup. Descriptive complexity on non-Polish spaces. In: 37th International Symposium on Theoretical Aspects of Computer Science,
               2020, LIPIcs,   v. 154,   p. 8:1--8:16, Schloss Dagstuhl - Leibniz-Zentrum f\"{u}r Informatik, 2020.

\bibitem {br} %[Br13]
M.
de Brecht. Quasi-Polish spaces.
\emph{Annals of pure and applied logic}, \textbf{164}, (2013), 356--381.

\bibitem{br1}
M. de Brecht,  A. Pauly, M.  Schr\"oder. Overt choice. Computability, DOI: 10.3233/COM-190253, arXiv 1902.05926v1 [math. LO], 2019.

\bibitem{he99} P. Hertling. An effective Riemann mapping theorem. Theoretical Computer Science, 219 (1999), 225--265.

\bibitem{hs}  M. Hoyrup,  C. Rojas,  V. Selivanov, D. Stull. Computability on quasi-Polish spaces. Proc. of DCFS-2019,  LNCS volume 11612, Berlin, Springer, 2019,   P. 171--183.

\bibitem{ke95}
A.~S. Kechris.  {\it Classical descriptive set theory}, Graduate Texts in Mathematics,
156, Springer, New York, 1995.

\bibitem{kk17}  M.V. Korovina, O.V. Kudinov.  On higher effective descriptive set theory. CiE-2017, LNCS volume 10307, Berlin,  Springer, pp. 282--291.

\bibitem{km19} T. Kihara, A. Montalb\'{a}n. On the structure of the Wadge degrees of bqo-valued Borel functions, Trans. Amer.
Math. Soc. 371 (2019), no. 11, 7885--7923.

\bibitem{lo78} A. Louveau. Recursivity and compactness. In: Muller, G.H., Scott, D.S.
(eds.) Higher Set Theory, Lecture Notes in Mathematics 669 (1978), pp. 303--337. Springer,
Heidelberg.

\bibitem{mo09} Y.N.  Moschovakis. {\em Descriptive  Set  Theory}.
North Holland, Amsterdam, 2009.

\bibitem{pa17} A. Pauly. Computability on the space of countable ordinals. Arxiv: 1501.00386v3, 2017. 

\bibitem{ro67}  H. Rogers, jr. {\em  Theory of Recursive Functions and
Effective Computability}. McGraw-Hill, New York, 1967.

%\bibitem{s82} Selivanov, V.L.:  On index sets in the Kleene-Mostowski hierarchy. {\em Trans. Inst. Math.}, Novosibirsk, N 2 (1982), 135--158 (Russian).

\bibitem{s83}  Selivanov V.L. Hierarchies  of
hyperarithmetical  sets and functions. {\em Algebra i Logika,} 22, No 6 (1983), p.666--692 (English translation: {\em Algebra and Logic,} 22 (1983), p.473--491).

\bibitem{s03}
V.L. Selivanov. Wadge degrees of $\omega$-languages of deterministic Turing machines.
{\em Theoretical Informatics and Applications}, 37 (2003), 67--83.

\bibitem{s06}
V.L. Selivanov. Towards a descriptive set theory for domain-like structures.
Theor. Comput. Sci. 365 (3), 258--282.

\bibitem{s08}
V.L. Selivanov. Fine hierarchies and $m$-reducibilities in theoretical computer
science. {\em  Theoretical Computer Science}, 405 (2008), 116--163.

%\bibitem{s08} Selivanov, V.L.: On the difference hierarchy in countably based
%$T_0$-spaces. {\em Electronic Notes  in Theoretical Computer Science}, V. 221(2008), 257--269.

\bibitem{s12} V.L. Selivanov. 
Fine hierarchies via Priestley duality. Annals of Pure and Applied
Logic, 163 (2012) 1075--1107.

\bibitem{s13} V.L. Selivanov.  Total representations. Logical Methods in Computer Science Vol. 9 (2013), pp. 1--30,
www.lmcs-online.org

\bibitem{s15} V.L. Selivanov. Towards the effective descriptive set theory. Proc. CiE 2015, LNCS volume 9136, Berlin, Springer, 2015, P. 324--333.

\bibitem{s17} V.L. Selivanov. Towards a descriptive theory of cb0-spaces. Mathematical Structures in Computer Science. v. 28 (2017), issue 8, 1553--1580.

\bibitem{s19} V.  Selivanov. $Q$-Wadge hierarchy in quasi-Polish spaces. Arxiv: 1911.02758v1, 2019.

\bibitem{sr07}
J. Saint Raymond. Preservation of the Borel class under countable-compact-covering mappings. Topology and its Applications 154 (2007), 1714--1725.

\bibitem{wei00} K.  Weihrauch. {\em Computable Analysis}. Berlin, Springer, 2000.

\bibitem{zi06} M. Ziegler. Effectively open real functions. Journal of Complexity, 22 (2006), 827--849.





\end{thebibliography}
\end{document}